\documentclass[pra,twocolumn,preprintnumbers,amsmath,amssymb,nofootinbib]{revtex4}
\usepackage{graphicx}
\usepackage[T1]{fontenc}
\usepackage[ansinew]{inputenc}
\usepackage{ae,aecompl}
\usepackage{enumerate,amsthm,amsmath,amssymb,color,float}

\begin{document}
\title{Violation of  Bell's inequalities in a quantum realistic framework.}
\author{Alexia Auff\`eves$^{(1)}$ and Philippe Grangier$^{(2)}$} 
\affiliation{
(1): Institut N\' eel, 25 rue des Martyrs$,\;$F38042 Grenoble Cedex 9$,\;$France \\
(2): Laboratoire Charles Fabry, 
%Institut d'Optique Graduate School, CNRS, Universit\'e Paris-Saclay, F91127 Palaiseau, France
IOGS, CNRS, Universit\'e Paris-Saclay, F91127 Palaiseau, France
 }
\begin{abstract}

We discuss the recently observed ``loophole free" violation of Bell's inequalities in the framework of  a physically realist view of quantum mechanics, which requires that physical properties are attributed jointly to a system, and to the context in which it is embedded. This approach  is clearly different from classical realism, but it does define a meaningful ``quantum realism" 
 from a general philosophical point of view. Consistently with Bell test experiments, this quantum realism embeds some form of non-locality, but does not contain any action at a distance,  in agreement with quantum mechanics. 

\end{abstract}
\maketitle

\section{Introduction}

A series of recent achievements  \cite{Delft,Vienna,Boulder} have convincingly demonstrated that Bell's Inequalities (BI) are violated, and that all previous ``loopholes" can be closed, provided that they are  experimentally testable  \cite{Bell-EPR,aaview}. One can thus conclude that  Bell's Hypotheses (BH), i.e. the physical and mathematical assumptions leading to BI, do not correspond to an acceptable description of nature. 

The precise implications of this statement remain open, especially if one asks about the resulting description of physical reality, as offered by Quantum Mechanics (QM). Quoting for instance Scott Aaronson \cite{Scott}, one would have to choose between ``to describe the ``reality" behind quantum processes via the Many-Worlds Interpretation, Bohmian mechanics, or following Bohr's Copenhagen Interpretation, refuse to discuss the ``reality'' at all''.

Here we want to move away from this apparent dilemma, by considering  that there is little to change to Bohr's Copenhagen Interpretation to obtain a fully consistent  ``quantum realism", compatible with QM and with the above experiments, but also with physical realism, defined as the statement that {\it the goal of physics is to study entities of the natural world, existing independently from any particular observer's perception, and obeying universal and intelligible rules.}\footnote{There is a very old but still alive philosophical debate   about what comes first: Are universal and intelligible rules a meaningful  idealization of empirical reality, which ought to be considered as the only reality? This approach can be called the Aristotelian point of view. Or are universal and intelligible rules the ultimately real, existing on their own in the realm of Plato's world of Forms? This is usually viewed as the Platonistic approach. We do not have to take a position in this debate, since in our definition of physical reality, the ``entities of the natural world" (Aristotelian reality) and  the ``universal and intelligible rules" (Platonistic reality)  are both needed. As physicists we
take for granted that the material world has an objective existence independent of observers, and that mathematical 
concepts are crucial to describe its properties. }

This {\bf quantum realism} has been presented in ref. \cite{FooP},  under the acronym CSM, standing for Contexts, Systems and Modalities. Here we will briefly summarize its main features, and discuss in more details how to use it to better understand the failure of BH. As it is well known, this failure of BH corresponds to a rejection of local realism, but not - as we will show - of physical realism. It can rather be considered as an evidence for a quantum realism, which is clearly different from classical realism, and which has some specific non-local feature - however, these features have nothing to do with any ``spooky action a distance".  
Generally speaking, the compatibility of QM with physical realism has been much debated in the literature \cite{EPR,EPR-wave,Bohr,Heisenberg,Bell,Moon,Contextuality,pg2,Mermin,PBR}, giving rise to many different interpretations of QM \cite{Frank}. 
In our approach  the quantum formalism and physical realism can perfectly coexist,  at the price of a subtle but deep change in  what is meant by physical properties: they are not any more considered as properties of the system itself, but jointly attributed to the system, and to the context in which it is embedded (definitions will be given below). We will show also that this ontological change has strong links with quantization as a basic physical phenomenon, and that this can explain why QM must be a probabilistic theory. 

This article is closely related to \cite{FooP}, with some parts condensed and others expanded, in order to spell out how the CSM approach explains quantum non-locality.

\section{System, context, and modalities}

To define an ontology  within the physical framework we are interested in, we will start with the question: which phenomena can we predict with certainty, and obtain repeatedly~?  Here certainty and repeatability of  
phenomena will be  used to provide necessary conditions to be able to define a ``state''. Such an approach, supported by quantum experiments, has a clear relationship with the criteria for physical reality given  in 1935 by Einstein, Podolsky and Rosen (EPR) \cite{EPR} -- but the ``object" to which it applies will be quite different \cite{Bohr}. 
%\\
%\\

Our quantum ontology involves three different entities. First comes the \textbf{system}, that is a subpart of the world that is  isolated well enough to be studied. The system is in contact with other systems, that can be a measuring device, an environment - no need to be more specific at this point. The ensemble of these other systems will be called a \textbf{context}. A given context corresponds to a given set of questions, that can be asked together to the system about its physical properties. A set of answers that can be predicted with certainty and obtained repeatedly within such a context will be called a \textbf{modality}.  

Given these definitions, let us bind them together by the following rule: {\bf In QM, modalities are attributed jointly to the system and the context. } This principle will be called ``CSM", referring to the combination of Context, System, and Modality.  As a set of certain and repeatable phenomena, a modality fulfills  the above conditions for the objective definition of a quantum state, and within the usual QM formalism (which is not here yet), a modality corresponds to a pure state. On the other hand,  the context is classical, in the sense that no other context has to be specified to define its state, and within the usual QM formalism, it corresponds to the parameters defining the observables as operators.  We note that here neither size, nor a quantitative criterium has been made to draw the quantum-classical boundary:  the quantum vs classical behavior is only related to the CSM principle itself, i.e., to the very definition of a modality. 

Taking a single polarized photon as an example\footnote{Interferences with polarized light (or in the quantum domain, with photon polarization) provides the simplest illustration of our approach. For interferences in the spatial domain,  it is more convenient to use also two-mode systems, like the Mach-Zehnder interferometer \cite{wheeler}, where the contexts are either (i)  the which-path detection within the interferometer (with two mutually exclusive modalities : upper path or  lower path), or (ii) the interferometer outcome detection. In quantum information, such a system is also known as a ``dual rail qubit".  Other interferometers like  Young's slits, or Fresnel's  biprism \cite{cachan}, are  more complicated to analyse because they are multimode systems, with ``fringes" where the interference phase depends on the position at the detection screen.  Then the outcome  can be analyzed in a larger Hilbert space; anyway the structure of  mutually exclusive modalities in a given context, and incompatible modalities between different contexts, is a very general framework for all quantum interference experiments.}, the system is the photon, the $\theta$-oriented polarizer is the context, and the two mutually exclusive modalities in this context are either ``transmitted", or ``reflected".  In the CSM perspective, a photon does not ``own" a polarization, but the ensemble photon-polarizer does.  If the context is known, and if the system is available, a modality defined in this same context can be recovered without error. This property has been exploited for years in quantum communications, and provides the core of quantum cryptography protocols \cite{BB84}. Here, we draw the consequences of this behavior in ontological terms. 
%\\

The resulting ontology is clearly different from the classical one,  where it is expected that a state should ``exist" independently of any context.
But even if CSM is fundamentally non-classical, physical realism is not lost: it still pertains to the ensemble made of context, system, and modality. Objectivity, defined as the independence from any particular observer's perception, is still guaranteed, but {\it the ``object" comprises both the system and the context, and its ``properties" are modalities \cite{Contextuality,pg2}}.

\section{Quantization and probabilities}

Now, a basic feature  is that in a given context, the modalities are ``mutually exclusive'', meaning that if one modality is true, the others are wrong. On the other hand, modalities obtained in different contexts are generally not mutually exclusive: they are said to be ``incompatible", meaning that if one modality is true, one cannot tell whether the others are true or wrong.
This terminology applies to modalities, not to contexts, that are classically defined: changing the context results from changing the measurement apparatus at the macroscopic level, that is, ``turning the knobs''. 
These definitions allow us to state the following quantization principle:

{\bf 
(i) For each well-defined system and context, there is a discrete number $N$ of mutually exclusive modalities; the value of $N$ is a property of the system within the set of all relevant contexts, but does not depend on any particular context. 

(ii) Modalities defined in different contexts are generally not mutually exclusive, and they are said to be ``incompatible''.}

Otherwise stated, whereas infinitely many questions can be asked, corresponding to all possible contexts, only a finite number $N$ of mutually exclusive modalities can be obtained in any of them\footnote{This principle is reminiscent of other approaches which bound the information extractable from a quantum system  \cite{Rovelli,AZ}. However, in the realist perspective we chose, quantization has not a purely informational character, but characterizes reality itself.}. 
An essential consequence is that it is impossible to get more details on a given system by combining several contexts, because this would create a new context with more than $N$ mutually exclusive modalities, contradicting the above quantization principle. As shown in \cite{FooP}, this makes that quantum mechanics must be a probabilistic theory, not due to any ``hidden variables", but due to the ontology of the theory.  Looking for instance at photon polarization, the number $N=2$ makes it  impossible to define a  (certain and repeatable) modality corresponding to  the photon being transmitted through a polarizer oriented at $0^{\circ}$, {\bf and} through a polarizer oriented at $45^{\circ}$, because then there would be 4 such modalities, in contradiction with $N=2$. 
Therefore the only relevant question to be answered by the theory is: given an initial modality  in context $C_1$, what is the {\it conditional probability} for obtaining another modality when the context is changed from $C_1$ to $C_2$ ? This probabilistic description is the unavoidable consequence of the impossibility to define a unique context making all modalities mutually exclusive, as it would be done in classical physics. It is therefore a joint consequence of the quantization and CSM principles, i.e. that modalities are quantized, and require a context to be defined\footnote{With different (non-ontological) approaches, many authors have emphasized the importance of contexts in QM, see e.g. \cite{c1,c2}.}. 
%\\

\section{About the EPR-Bell argument}

\subsection{The  EPR-Bohm argument}

We can now discuss in more details the EPR argument \cite{EPR}\footnote{Here we consider the EPR-Bohm argument with spin 1/2 particles, rather than the original EPR argument with wave functions, which has some specific features discussed elsewhere \cite{EPR-wave}.} and Bell's theorem  \cite{aaview,Bell-EPR}. To do so,  let us consider two spin 1/2 particles in the singlet state, shared between Alice and Bob. The singlet state is a modality among four mutually exclusive modalities defined in a context relevant for the two spins, where measurements of the total spin 
(and any component of this spin)  will certainly and repeatedly give a zero value. On the other hand, the singlet state 
is incompatible with any modality attributing definite values to the spin components of the two separate particles in their own (spatially separated) contexts.
According to the previous section, the singlet modality is thus certain and repeatable in its own context (e.g., measurement of the total spin), but can only provide probabilities for the values of the spin components of the two separate particles. 

\subsection{What happens on Bob's side ? }
Now, let us assume that Alice performs a measurement on her particle, far from Bob's particle. Alice's result is random as expected, but what happens on Bob's side? Since Bob's particle is far away, the answer is simply that nothing happens. How to explain the strong correlation between measurements on the two particles? By the fact that after her measurement, Alice can predict with certainty the state of Bob's particle; however, this certainty applies jointly to the new context (owned by Alice) and to the new system (owned by Bob). The so-called ``quantum non-locality" arises from this separation, and the hidden variables from the impossible attempt to attribute properties to Bob's particle only, whereas properties must be attributed jointly to Alice's context and Bob's system. Getting them together is required for any further step, hence the irrelevance of any influence on Bob's system following Alice's measurement. Here the separation between context and system is particularly obvious and crucial, since they are in different places. 

\subsection{What and where is the ``reality" ?}
According to the above reasoning, after Alice's measurement on one particle from a pair of particles in a singlet state, the ``reality" is a modality for Bob's particle, within Alice's context.  But Bob may also do a measurement, independently from Alice, and then  the ``reality" will be a modality for Alice's particle, within Bob's context.   Does that mean that we have two ``contradictories" realities ?  Actually no, because these realities are contextual \cite{Contextuality,pg2}:  for instance Alice's modality tells that if Bob uses the same context as Alice, he will find with certainty a result opposite to Alice's one (given the initial singlet state). This statement is obviously true, as well as the one obtained by exchanging Alice and Bob. 
But if Bob does a measurement in another context (different from Alice's), then one gets  a probabilistic change of context for a $N=2$ system,  as described before.

If Alice and Bob both do measurements with different orientations of their analyzers, the simplest reasoning is to consider 
the complete context for both particles,  which is initially a joint context (with a modality being the singlet state) and finally two separated contexts, 
again with 4 possible modalities due to the quantization postulate. Then this is now a probabilistic change of context for a $N=4$ system,  again with the same result. 

\subsection{Where does CSM differ from Bell's hypothesis ?}
It is interesting to write a few equations about these initial,  ``intermediate" and final  modalities,  
because this allows us to see more explicitly where CSM differs from Bell's hypothesis, even before the quantum formalism is introduced. 
So let us denote $a_i$, $b_j$  the modalities with results $i, \; j = \pm 1$ for some orientation (context)  $a$ for Alice, and $b$ for Bob. Given some ``hidden variables" $\lambda$, and using the vertical bar ``|"  as the usual notation for conditional probabilities $p(X|Y)$, the core of Bell's hypothesis is to assume the factorisability condition :
\begin{equation}
p(a_i, b_j | \lambda) = p(a_i | \lambda) \; p(b_j | \lambda)  \label{bellhyp}
\end{equation}
The equivalent CSM equations, given the initial joint modality $\mu$, are for Alice,  who knows $\mu$ and $a_i$
\begin{equation}
p(a_i, b_j | \mu) = p(a_i | \mu) \; p(b_j | \mu, a_i) \label{csmalice}
\end{equation}
whereas they are for Bob, who knows $\mu$ and $b_j$
\begin{equation}
p(a_i, b_j | \mu) = p(a_i | \mu, b_j) \; p(b_j | \mu).  \label{csmbob}
\end{equation}
It is clear that Eqs. (\ref{csmalice}, \ref{csmbob}) differ from Bell's hypothesis Eq. (\ref{bellhyp}), and therefore 
Bell's inequalities can be violated in the CSM framework, without requiring any
action at a distance, or faster than light signalling. However, there is some non-locality, in the sense that the result on one side depends on the result on the other side; but this is only through a (local) redefinition of the context, not through any influence at a distance onto the remote particle. Again, it is essential  to consider that the modality belongs jointly to the particle(s) {\bf and} to the context, and not to the particle(s) only, otherwise one would be lead to Bell's hypothesis.\footnote{This argument shines light on the recent animated exchange between Tim Maudlin and Reinhard Werner, see for instance arXiv:1408.1826, arXiv:1408.1828, arXiv:1411.2120, https://tjoresearchnotes.wordpress.com/2013/05/13/guest-post-on-bohmian-mechanics-by-reinhard-f-werner/. 
\\
In this rich discussion  it is correctly pointed out that an essential hypothesis of ``classicality" (or classical realism)  is embedded in Bell's hypothesis.  But according to CSM, it is not correct to claim that removing this hypothesis, and moving to some form of quantum realism,  should eliminate all problems with locality. Everything will be fine as far as relativistic causality is concerned, but a fully non-classical form of non-locality will remain, due to the fact  that the context and the system can be in different places. This  would make no sense classically, but it is essential  in a quantum framework, and explains why Bell's hypotheses do not hold.  The CSM approach  is the best way we know to spell out this specifically quantum non-locality, which does not imply any ``spooky action at a distance".}

\subsection{Does CSM agree with QM, and why ?}
Another important consequence is that  if Alice and Bob both do measurements, their realities must ultimately agree together, since there will be a unique final modality $(a_i, b_j)$. Therefore their predictions must also agree together, and one must have 
$$p(a_i, b_j | \mu) = p(a_i | \mu) \; p(b_j | \mu, a_i)  = p(a_i | \mu, b_j) \; p(b_j | \mu)   $$
These equations are just the same as the ones we would obtain by the usual ``instantaneous reduction of the wave packet", though in our reasoning there is no wave packet, and no reduction, but only a measurement performed by either Alice or Bob on the known initial modality $\mu$. Even more, if we admit that $(\mu, a_i)$ is a new modality for Bob, and $(\mu, b_j)$  is a new modality for Alice, then $p(b_j | \mu, a_i)$  or  $p(a_i | \mu, b_j)$  cannot be anything else than the one-particle conditional probabilities; for instance, it will be the usual   Malus law for polarized photons. 

Finally, it is worth emphasizing that from a physical point of view, the modality $ (\mu, a_i) $ obtained after Alice's measurement on the entangled state is exactly the same as the one that would be obtained by transmitting  a single particle in this same modality from Alice to Bob. This equivalence between an entanglement scheme and a ``prepare and measure" scheme has been extensively used in security proofs of quantum cryptography. 

So we get a simple explanation about  the famous ``peaceful coexistence" between QM and relativity, i.e. why quantum correlations are non-local, but also ``no signalling" (they  don't allow one to transmit any faster than light signal): this is because when Alice makes a measurement, the change from 
 $\mu$ to  $(\mu, a_i)$ corresponds to a change of context, and not to any influence at a distance. This change of context (from joint to separate) redefines a new modality, which always involve both a system and a context. Such a situation, though strongly non-classical, does not conflict with physical realism  or causality: in the CSM perspective, quantum non-locality is a direct consequence of the  bipartite nature of quantum reality.

\section{Conclusion}

Beyond the previous discussion on the EPR-Bell argument, CSM can answer  several ontological questions posed by QM. In what follows, we  conclude this paper with a short review of some other topics that this new perspective can help clarify. As this is being done, realism is again asserted, and objectivity is maintained,  when applied to contexts, systems and modalities.
\\

\noindent {\it  Quantum realism and ontology.} 
Contextual objectivity \cite{Contextuality,pg2} allows for a quantum ontology, as the joint reality of the context, system, and modalities (CSM). This allows us to interpret quantum nonlocality as the situation where the context and the system are separated in space.
Such a situation has no conflict with physical realism, but is irrelevant in classical physics, where the physical properties are carried by the system alone. 
\\

\noindent {\it  Accepting the shifty split.} 
For many physicists, putting the context at the very heart of the theory implies an unacceptable ``shifty split" \cite{Bell,Mermin} between the quantum world (of the system) and the classical world (of the context). A lot of efforts have been made to get rid of it, and to make the classical world emerge from the quantum world, by attempts to describe  contexts within the quantum formalism. Such attempts may exploit the fact that there is a lot of  flexibility for defining the boundaries of the system, especially when considering that (weak or strong) measurements can be done by entangling the initial system with more and more ``ancillas", leading to the so-called ``Von Neumann regression" \cite{JvN}. 

But in our approach, extending measurements to include the context is self-contradictory: even by adding many ancillas, 
the system can never grow up to the point of including the context, simply because without the context, modalities cannot be defined. 
In other words, looking at  the system as a fuzzy object including everything is not consistent with our physically  realist ontology. 
The quantum-classical boundary has therefore a fundamental character, both from a physical and from a philosophical point of view \cite{FooP}.  Without restricting the generality nor the applicability of QM, the CSM approach acknowledges  that, as a scientific discipline, QM ``can explain anything, but not everything" \cite{Peres-Zurek}.
\\

\noindent {\it  The Copenhagen point of view.} 
In its practical consequences CSM is close to the usual Copenhagen point of view (CPV), so it may be interesting to discuss also the differences.  A crucial one is that quantum reality as defined in CSM deviates from CPV, where reality is rather a word to be avoided \cite{Scott}.  Whereas CPV may be accused of dogmatism (hidden behind mathematical formulas), the ontological claims of CSM have some flavor of empiricism, or phenomenology: their goal is to provide a physically realist view of QM ``as it is done", including in all the recent BI tests. 
\\

\noindent {\it  Decoherence theory.} 
The practical side of CSM vs CPV can also be illustrated by considering ``decoherence theory" \cite{Zurek,PhT}.
Considering that in an actual measurement, the system interacts with ancillas, entanglement is created, and observations are made, decoherence theory provides criteria to decide when and why a ``big" ancilla does not behave as a quantum system any more. But  this is done by using QM, and thus - in the CSM view - this only makes sense with respect to an external context, always required for defining modalities and using the quantum formalism \cite{PhT}. Said otherwise, starting from a vector $|\psi \rangle$ in an Hilbert space, and then trying to ``deduce" the classical world,  appears as circular by construction, because (from the beginning) $|\psi \rangle$ is a mathematical object associated with a modality, i.e. with a phenomenon involving both the ``classical" and ``quantum" worlds. Therefore decoherence theory perfectly fits  within CSM, being admitted that the goal is not to reconstruct the classical world (it is already there) but  to show  that QM  is a consistent theory. Said otherwise, QM is extraordinarily efficient for managing the ``split", but cannot get rid of it, because it is built within the quantum ontology and expressed in the quantum formalism - loosely speaking, as the difference between observables (contexts) and states (modalities). 
%\\
\vskip 2mm

\noindent {\it  The Bohr-Einstein debate.} 
As a final remark, Bohr's arguments in \cite{Bohr} were quite right, but perhaps failed to answer a major question asked in essence by EPR in  \cite{EPR}: can a physical theory be ``complete" if it does not provide an ontology that should be clearly compatible with physical realism~? Unveiling such a realistic quantum ontology is what is proposed by our approach. 
%\\
\vskip 2mm

%\subsubsection*{Acknowledgements}
\noindent{\bf Acknowledgements:} 
The authors thank Nayla Farouki for essential contributions, and Franck Lalo\"e, Francois Dubois, Anthony Leverrier, 
Maxime Richard,  Augustin Baas, Cyril Branciard for many useful discussions. 

%\vskip 1.5cm

\end{document}